\documentclass[sigconf]{acmart} 
\AtBeginDocument{%
  }

\setcopyright{acmlicensed} 
\copyrightyear{2018} 
\acmYear{2018} 
\acmDOI{XXXXXXX.XXXXXXX} 
\acmConference[Conference acronym 'XX]{Make sure to enter the correct
  conference title from your rights confirmation email}{June 03--05,
  2018}{Woodstock, NY}  
\acmISBN{978-1-4503-XXXX-X/2018/06}  

\usepackage{amsmath,amsfonts}
\usepackage{algorithmic}
\usepackage{graphicx}
\usepackage{textcomp}
\usepackage{xcolor}
\usepackage{pifont}
\usepackage{tikz}
\usepackage{url}

\usetikzlibrary{positioning, arrows.meta, calc}

\begin{document}


\title{Compile-time Security Analysis and Optimization of Sensitive String Producers} 



\author{Mike Samuel}
\affiliation{%
  \institution{Temper Systems, Inc.}
  \city{Denver}
  \state{CO}
  \country{USA}
}
\email{mikesamuel@gmail.com}
\orcid{0009-0006-9271-1732}

\author{Tom Palmer}
\affiliation{%
  \institution{Temper Systems, Inc.}
  \city{Tucson}
  \state{AZ}
  \country{USA}
}
\email{tom@temper.systems}
\orcid{0000-0002-1593-0011}

\author{Shaw Summa}
\affiliation{%
  \institution{Temper Systems, Inc.}
  \city{Scranton}
  \state{PA}
  \country{USA}
}
\email{shawsumma@temper.systems}
\orcid{0009-0004-9018-9229}

\author{Robert Grayson}
\affiliation{%
  \city{New York}
  \state{NY}
  \country{USA}
}
\email{bobbbygrayson@gmail.com}
\orcid{0009-0008-9990-5526}

\renewcommand{\shortauthors}{Grayson, Palmer, Samuel and Summa}

\begin{abstract} 
Content composition vulnerabilities remain among the most prevalent and persistent classes of security weakness in deployed software. Prior mitigations, including developer training, static analysis tools, and domain-specific template languages, each face diminishing returns; AI code generation inherits these limitations and introduces new ones, reproducing insecure patterns from training data and lacking reliable context for self-correction.

This paper introduces a general framework for secure content composition that extends across content languages and integrates directly into general-purpose programming languages via additive changes to string expression syntax.  We define a language design goal of minimizing the lexical distance between secure and insecure idioms, and show that this goal admits practical compilation strategies: static analyses specified in terms of dynamic semantics, runtime performance approaching na\"ive string concatenation, and developer-facing diagnostics surfaced as compile-time errors or warnings.

The approach enables an effective division of labor: security engineers encode composition hazards in libraries once; developers and AI coding agents select the appropriate library primitive to implement features correctly without needing to internalize specialist security knowledge; compiler diagnostics provide objective, position-keyed feedback that grounds both human review and iterative AI self-correction; and security responders focus on keeping libraries current rather than auditing ad-hoc security decisions distributed across a codebase.
\end{abstract}




\maketitle

\section{Background}
\label{sec:background}

Many of the most persistent classes of computer security vulnerability are \emph{content composition} problems\thinspace\cite{cwe-top25-2025}: XSS, SQL-injection, command injection, code injection, and path traversal attacks. They arise when a program assembles a string in a complex structured language, e.g. HTML, SQL, JSON, file path syntax, or Bash, from a mixture of trusted, fixed parts and untrusted, dynamic expression results, and the resulting string is then wholly trusted by a downstream consumer such as a browser, database engine, web API server, or operating system.

Specialist security engineers understand how to compose such content safely.  The key insight is that the \emph{fixed, trusted parts} of a content string carry implicit context; they determine which syntactic position each \emph{dynamic, untrusted expression result} occupies in the output language's grammar.  Knowing that position determines the correct escaping function to apply to each dynamic value, rendering the composed string trustworthy as a whole.  Encoding this knowledge correctly requires careful study of the target content language; getting it wrong silently produces insecure output.

The challenge of secure content composition is not new, but the emergence of AI-assisted code generation gives it renewed urgency and suggests an avenue that has received little attention: adapting programming languages themselves so that they become better targets for AI code generation. Static analysis tools (SAST), developer security training, and code review have each reduced vulnerability rates, but each hits diminishing returns as the low-hanging fruit is addressed or the threat environment changes faster than training materials can disseminate or training corpora incorporate new best practices.  AI code generation inherits these limitations: models trained on existing code reproduce the insecure patterns prevalent in that code, and prompting techniques such as Recursive Criticism and Improvement (RCI)\thinspace\cite{kim2023rci} that iteratively refine generated code for security can improve outcomes, but are ultimately bounded by the model's ability to identify its own mistakes in the absence of targeted feedback. A more durable approach is to make secure by construction the path of least resistance: when the secure idiom is syntactically close to the insecure one, and when the compiler surfaces actionable diagnostics, both human developers and AI agents have the tools they need to produce secure systems.

This paper argues that language design is an underexplored partner in the effort to improve the security of both human-written and AI-generated code, and that the class of content composition vulnerabilities is a particularly tractable starting point. More importantly, a general framework for this class gives a pathway to addressing new instances as code generation changes the way software is developed and as new composition hazards (perhaps heretofore unseen injection vulnerabilities in agentic systems) replace existing ones as the most critical to address. The specific approach outlined here, contextual autoescaping aided by specific API design and type alignment choices, generalizes an approach that has been deployed at scale to great effect\thinspace\cite{google2024securebydesign}.

\section{Introduction}
The first section discusses string composition constructs in modern programming languages, and outlines goals for programming language idioms that support generalist programmers in safely composing sensitive strings using libraries written by specialist security engineers.

The second section discusses how specialist security engineers can encode analysis of common wire formats like HTML with special attention to language nesting, escaping, and attack-surface reduction techniques.  It introduces \textit{transition tables} as a way of specifying automata for various content languages, and explains how automata for one language can delegate to automata for nested languages.

The third section talks about compiler techniques for performing, at compile time, the analysis that library code specifies in terms of runtime code.  This section talks about predictive execution of pure function calls, but is applicable to programming languages that include impure functions.

The last section relates this to prior work including idioms that embed support for one content language in a general purpose programming language or auto-escaping, domain-specific template languages.

\section{String Composition Syntax}
\label{sec:string-composition-syntax}

\subsection{Readability and Familiar Syntax}

Readable string composition syntax is not merely an ergonomic convenience; it directly affects the ability of maintainers and security auditors to comprehend the structure of the composed strings, and confidently relate that to higher-level goals.  Figure~\ref{fig:kotlin-vs-java} contrasts two code fragments that construct strings: a Kotlin quoted expression and Java using \texttt{StringBuilder}.

\begin{figure}[htbp]
\begin{verbatim}
message = "Hello, ${whom ?: "World"}!"
\end{verbatim}
\hrule
\medskip
\small
\begin{verbatim}
var sb = new StringBuilder();
sb.append("Hello, ");
if (whom != null) {
  sb.append(whom);
} else {
  sb.append("World");
}
sb.append("!");
message = sb.toString();
\end{verbatim}
\caption{Kotlin string template (top) versus Java \texttt{StringBuilder}
  (bottom).  The Kotlin form makes the structure of the output string
  immediately apparent; the Java form requires the reader to mentally
  reassemble the output from sequential imperative steps.}
\label{fig:kotlin-vs-java}
\Description[Two code blocks side by side]{The top is one line of Kotlin that constructs a string and assigns it to message, using interpolation syntax with an infix ?: fallback to the string "World" when the expression is null.  The bottom is Java code that does the same but involves nine lines of code the first of which creates a StringBuilder.}
\end{figure}

The trend in modern languages is toward multi-line strings, and quoted expressions with \emph{inline expressions}. Java~JEP-378\thinspace\cite{jep378} adds multi-line text blocks to the Java language. Python~PEP-498\thinspace\cite{pep498}. introduced "f-strings" despite the language already offering sprintf-style format-string syntax such as \verb|"%s, %s!" % (greeting, whom)|. There's a downside to format strings: when multiple values are interpolated, format-string syntax requires the reader's eye to traverse back and forth between placeholder positions and argument positions. Inline interpolation syntax eliminates this indirection by placing each expression at its point of use in the output structure.

\subsection{Expressive Power: Control Flow and Embedded Statements}

The PHP language demonstrates that embedding conditionals and loops directly inside content-generating code affords substantial flexibility, allowing business logic to govern branching without obscuring the high-level structure of the output. The remainder of this paper assumes a quoted expression syntax that supports four structural elements:

\begin{enumerate}
  \item a \emph{tag} which controls how the expression result is composed;
  \item runs of fixed, \emph{literal characters};
  \item interpolated, dynamic \emph{expressions}; and
  \item \emph{embedded statement fragments} which allow control flow (\texttt{if}, \texttt{for}) to govern which parts are emitted.
\end{enumerate}

The example syntax in Figure~\ref{fig:tagged-sugar} desugars to the code shown in Figure~\ref{fig:tagged-desugar}. An \textit{accumulator} is an object that knows how to compose fixed, trusted parts and untrusted interpolations. It typically owns a \textit{collector}, like the \textit{StringBuilder} from Figure~\ref{fig:kotlin-vs-java}, onto which it appends content after appropriate checks and transformations. This separation between accumulators and collectors enables compile-time erasure discussed in section \ref{subsec:accumulator-erasure}.

\begin{figure}[htbp]
\begin{verbatim}
let title = "Count", verb = "count";
// Start at 0. I am a vampire, not a monster.
let items = ["Zero", "One", "Two"];

let drama = cackle"""
  "I am the ${title} who loves to ${verb}!
  :for (let item of items) {
    "${item}! Ha ha ha.
  :}
  "Brought to you by the number ${items.last()}.
\end{verbatim}
\caption{A tagged block expression using the \texttt{cackle} tag.
  Lines whose leftmost margin character is \texttt{"} contribute
  literal runs and interpolations to the output; lines whose leftmost
  margin character is \texttt{:} contribute balanced statement
  fragments that govern control flow.}
\Description[A block of Temper code]{The code declares and assigns some local names and then has a complex string expression with a tag and with an embedded for loop.}
\label{fig:tagged-sugar}
\end{figure}

\begin{figure}[htbp]
\small
\begin{verbatim}
let tag = cackle;
let accumulator = tag.newAccumulator();
accumulator.appendFixed("I am the ");
accumulator.appendUnsafe(title);
accumulator.appendFixed(" who loves to ");
accumulator.appendUnsafe(verb);
accumulator.appendFixed("!\n");
for (let item of items) {
  accumulator.appendUnsafe(item);
  accumulator.appendFixed("! Ha ha ha.\n");
}
accumulator.appendFixed(
  "Brought to you by the number ");
accumulator.appendUnsafe(items.last());
accumulator.appendFixed(".\n");
drama = accumulator.collected();
\end{verbatim}
\caption{Desugaring of the tagged block expression in
  Figure~\ref{fig:tagged-sugar}.  Fixed parts are passed to
  \texttt{appendFixed}; dynamic expression results to
  \texttt{appendUnsafe}.
  The tag's \texttt{collected} call produces the final value.}
\label{fig:tagged-desugar}
\Description[A block of Temper code]{The syntactic desugaring of the previous figure.}
\end{figure}

After a triple-quote delimiter (\texttt{"""}) each line of the expression starts with a margin character: a quote (\texttt{"}) for lines of literal characters and interpolations, or a colon (\texttt{:}) for lines of statement tokens, which taken together, must form a balanced statement.

\subsection{Robustness Under Maintenance}

A key motivation for supporting embedded control flow is practical robustness over a codebase's lifetime.  Quoted expressions frequently begin as simple, interpolation-only templates, but as maintainers address edge cases (handling a null value differently, omitting an optional section, iterating over a collection) they require additional expressiveness.  A secure composition idiom that covers only the simple case will be abandoned in favor of insecure alternatives when changing requirements call for a conditional or a loop.  Allowing embedding control flow within the same syntactic construct ensures that the secure idiom remains the path of least resistance as developers work to address non-security-related bugs and feature requests.

\subsection{Front-Loading Context for Readers and Auditors}

The tagged quoted expression syntax has a further property that helps both human code reviewers and AI auditors agents: it front-loads the information most relevant to understanding the desired properties of the output.  A reader encountering \texttt{let drama = cackle"""} immediately learns the destination name (\texttt{drama}) and the tag (\texttt{cackle}), and can infer the type of the composed value, before reading any of the content.  By contrast, the Java \textit{StringBuilder} code from Figure~\ref{fig:kotlin-vs-java}, reveals the destination, \texttt{message}, only after the reader has processed every accumulator call. Similarly \texttt{httpResponse.bodyContent = html"""} helps the reader adopt an appropriate frame of mind for what follows.

The end result of these language design decisions is that the lexical difference between secure and insecure idioms is small, but communicates intent clearly.  A code auditor inspecting a quoted expression can immediately identify the output language or the lack of a tag that specifies it.  Where the output language matters for security, a reviewer can suggest a small, actionable change, adding or replacing a string tag, rather than a refactoring that touches multiple call sites.  This property is particularly valuable for \emph{adversarial review}. When the syntactic structure of an expression mirrors the structure of its output, it is easier to reason about whether the output has the right property, and harder for an ill-conceived or malicious patch to hide structural violations within a soup of imperative statements.

AI models do not need to embed the expert knowledge of specialist security engineers if they can delegate security critical tasks to libraries written by those engineers.  Current models are already capable of selecting the right library primitive for a task when appropriate primitives exist and are findable\thinspace\cite{kim2023rci}.  The language design goal pursued here, making the secure idiom the obvious choice at the point of use, is precisely the kind of signal that allows a code-generator to make the correct choice without requiring it to reason from first principles about HTML escaping or SQL quoting rules.  There is a practical efficiency argument here as well: AI models have limited attention windows and human developers rarely read verbose documentation.  Short, simple rules of thumb of the form \emph{``use the \texttt{html} tag when producing HTML''} serve both audiences better than reams of content-language specific rules like \emph{``when interpolating URL attribute values into HTML, apply an escaper that does X, Y, and Z''}.

\section{Implementing contextual auto-escaping}
\label{sec:implementing}

\subsection{Runtime Semantics via the Accumulator Interface}

A library author who wishes to define a new content type specifies its runtime composition semantics by providing an \emph{accumulator} type that receives two kinds of input from the compiler-generated desugaring: calls to \texttt{appendFixed}, which deliver chunks of literal characters drawn directly from the program source and therefore trusted, and calls to \texttt{appendUnsafe}, which deliver the results of interpolated expressions and are therefore untrusted.  The tag library is then free to implement whatever context-tracking, escaping, or rejection logic it requires inside these two operations, producing a final value of the library's content type via \texttt{fromCollector}.

The compiler does not need to understand the content language at all. The accumulator type's semantics fully specifies how string expression parts compose to form the final result.

\subsection{Purity and Optimization}

If certain of the accumulator type's operations are implemented in a \emph{pure, side-effect-free subset} of the host language, or as appends do, have a separable effect limited to appending to a list with a scoped lifetime, the compiler gains important latitude for optimization.  Because pure functions have no observable side effects, the compiler can freely reorder, inline, or speculatively evaluate calls without risk of changing program behavior.  This also simplifies reasoning for both the compiler and human reviewers: a pure accumulator implementation can be analyzed in isolation, without considering interactions with the rest of the program's state, and is easier for formal methods practitioners to reason about. A language's compiler may reserve additional compiler passes for accumulator implementations whose key operators fall into this subset.

Joe-E\thinspace\cite{finifter2008joe-e}, a dialect of Java, is an impure language with a declared, pure subset. The Joe-E language curates global names to restrict ``ambient authority''; e.g. it denies access to \textit{java\allowbreak .lang\allowbreak .Runtime}'s \textit{exit(int)} method which has a side effect. Select Joe-E methods can be declared effect-free and Joe-E types can be declared deeply immutable.  The Joe-E compiler restricts name bindings in pure methods, allowing imperative effects on local variables and mutable book-keeping objects created by the method call to compute the result, but maintaining the property that each call has no externally visible effects and its output does not depend on how it interleaves with operations that do have visible effects.

The ability to reorder some calls enables speculative execution of method calls that compute results of security analyses. This enables the language toolchain to shift dynamic analysis for security from runtime to compile time where a strict response to one problem is unlikely to result in a denial of service.

\subsection{Left-to-Right Parsing and Context Propagation}
\label{subsec:lrparsing}

The key requirement that a tag library must satisfy is the ability to parse its content language \emph{purely from left to right}, in a single pass, suspending and resuming at interpolation boundaries.  At each such boundary the parser holds a \emph{context}: a value that summarizes everything about the parse state accumulated so far.  That context determines two things: first, the \emph{escaper} to apply to the upcoming untrusted expression result, so that the result is safe to embed at this position in the content language; and second, the \emph{successor context} that follows the escaped value, from which parsing of the next fixed chunk will continue.

The fundamental unit of analysis consists of a context, an escaper derived from it, and a transition to a next context.  The rest of this section explains how it can be specified in practice.

\subsection{Automata-Based Parsing: TextMate Grammars as a Model}
\label{subsec:textmate}

A practical question is what formalism to use for specifying the left-to-right context transitions. TextMate grammars\thinspace\cite{tarvankrieken-2023} have been successfully developed and maintained by programming language enthusiasts to encode language specific code syntax highlighting for Visual Studio Code and other code editors.

It uses an automaton-based approach that transitions when a regular expression successfully consumes a prefix of an input, and the transition controls which rules are applicable in future transitions. The order of rules matters, so adopting a similar approach allows security engineers to specify rules for dangerous corner cases first, and know that no regular expression overlap will lead to corner-case capture.

\subsection{Implementation goals}
\label{subsec:implementationgoals}

Writing a secure composition parser is not a straightforward mapping of syntactic elements to grammars because it involves reasoning about both parsing (context propagation) and unparsing (how untrusted interpolations will map to content substrings in the worst case), and in practice, there are often buggy parsers in the field. To guide implementors, here are some principles to help security engineers reason about properties of a high-quality composition library.

The \textit{structure preservation property} holds for a composition library when, regardless of the untrusted interpolations in a sequence, the fixed strings will cause the downstream parser to produce the same parse tree additions. Reasoning about a left to right parse with holes requires a few tricks, but thinking about concrete syntax tree (CST) element starts and ends independently instead of whole CST nodes can help.  A template like \verb|html"<a href=${...}>| has these effects on the parser when the template hole is filled with a valid, innocuous value: start a tag \#1, fill its name slot, start an attribute name \#2, fill its name slot, a dynamic attribute value, and finally the \texttt{>} ends tag \#1. If the interpolation could contribute characters like "\verb|><script>doEvil()</script|" then that first injected '\verb|>|' would break the relationship between the \verb|<a| starting tag\#1's eventual CST node and verb|>| ending that same tag: a violation of structure preservation.

The \textit{code effect property} holds when values in interpolations that are not privileged (as by satisfying a Morris\thinspace\cite{morris-1973} style trademark) cannot cause high-privilege instructions to execute, and cannot cause intended privileged instructions not to execute.  What constitutes a privileged value or a high-privilege instruction depends on the content language, but this topic is thoroughly covered in content languages' attack vectors literature. As an example, initiating a network request with the current user's cookies is a privileged effect, so injecting the plain string value \texttt{"javascript:\allowbreak sendMy\allowbreak CryptoWallet\allowbreak ToAlice()"} into an HTML URL attribute is clearly a violation. The latter rule about desired instructions not executing is meant to capture denial of service problems which attackers might trigger by introducing minor parsing errors. For example, JSON parsers are often strict in rejecting trailing commas, so if an interpolation can contribute zero characters where an array list item is expected, \texttt{["foo", ]}, then it can cause the larger JSON document to be wholly rejected.

Finally, Raymond\cite{raymond-taoup-2003}'s \textit{principle of least surprise} holds no less for content composition than for any other API or language design task. Code authors, reviewers and auditors need to confidently form correct conclusions about the meaning of code. This is very much a human factors concern, but warrants no less attention from security engineers than the other properties. In a later section, this paper talks about inlining escapers into call sites. Done carefully, this can allow integrated development environments (IDEs) to pop up escapers' documentation when the developer hovers over an interpolation boundary token, making escaping decisions more transparent and helping mitigate least-surprise violations.

\subsection{Epsilon Transitions, Substitutions, and Attack Surface Reduction}
\label{subsec:epsilon}

Epsilon transitions, parse state transitions that consume no input, are a source of complexity in parsers and unparsers, and also warrant special attention in secure composition schemes.  Consider an HTML fragment such as \texttt{<a href=\$\{x\} other-attr= \$\{y\}>}.

If the interpolated value of \texttt{x} contributes no characters, i.e. if the transition on the interpolation is an epsilon transition, then, from the perspective of the browser that receives this content, the \texttt{href} attribute value is immediately followed by \verb|other-attr=|, making the value of \texttt{href} appear to be the string \texttt{other-attr=} rather than nothing, and potentially allowing the interpolation of \verb|${y}| to specify both an unintended attribute name, and its value.  The author almost certainly intended an empty attribute value represented by the non-empty token sequence \verb|""|; the epsilon transition allows a hazardous misparse.

The remedy is to allow automaton transitions to specify an optional \emph{substitution}: a fixed string that is inserted into the output on the transition, independently of the interpolated value.  For the unquoted-attribute case, the automaton can insert quotation marks around the interpolation site, ensuring that the attribute value is correctly delimited regardless of whether the expression result is empty.  This also eliminates a historical class of vulnerability: Microsoft browsers interpreted non-ASCII typographic quote characters as attribute-value delimiters to accommodate early, novice web developers who started off writing HTML in MS WordPad without first disabling smart-quote substitution. The ability for an automaton to adjust a fixed portion reduces the attack surface by removing from the output language strings that might have variant interpretations on non-conforming parsers.

Substitutions by nested languages ought not interfere with structure preservation in the nesting language. For example in \texttt{html`<div style="background-image: url(\$\{x\})`></div>} the HTML embeds CSS which might need its \verb|url(...)}| syntax to be quoted, so the CSS automaton inserts quotes which then need to be HTML escaped to preserve the integrity of the HTML quotes, so after insertion, that is equivalent to \texttt{html`<div style="background-image: url(\&quot;\$\{x\}\&quot;)"></div>`}. Each subsidiary automaton is paired with a codec (coder, decoder) that expresses how the outer language can decode fixed text substrings before forwarding to a nested language's automaton, and how to re-encode any substitutions by the automaton which might contain outer-language meta-characters.

\subsection{Implementation: Transition Tables for the Web Stack}
\label{subsec:transition-tables}

This scheme has been implemented for the web stack content languages: HTML, CSS, JavaScript, and URLs. The library defines automata via explicit \emph{transition tables}. A transition table row specifies:

\begin{enumerate}
  \item a \emph{pattern} that matches on the current context's fields;
  \item a \emph{regular expression} that must consume a prefix
        of the current fixed string chunk, or an indicator that an
        interpolated value is arriving, or neither;
  \item an optional \emph{substitution} to emit in place of any matched prefix;
  \item a \emph{successor context} derived from the current context,
        together with optional adjustments to a subsidiary automaton
        (see below); and
  \item an optional \emph{message} to display when we know a compiler or developer's IDE is in a position to present errors and warnings.
\end{enumerate}

Figure~\ref{fig:transition-rules} shows a representative excerpt of
the HTML transition rules.  An HTML context consists of four state variables: a high level state, a tag classification, an attribute classification, and a kind of attribute delimiter.  This represents a high-level context, and is meant to be easily merged across control flow paths during static analysis though merging is not required during dynamic analysis.  It intentionally does not represent the entire path through an HTML grammar taken by a parser for the prefix seen thus far.

\begin{figure*}[htbp]
\small
\begin{verbatim}
| Pcdata, _, _, _ | `<\/(?=[a-zA-Z])`        |        | CName, _, _, _          |
| Pcdata, _, _, _ | `(?i)<script(?!{{nmc}})` |        | BeforeAttr,Script, _, _ |
| Pcdata, _, _, _ | `(?i)<style(?!{{nmc}})`  |        | BeforeAttr,Style,  _, _ |
| Pcdata, _, _, _ | `<(?=[a-zA-Z])`          |        | OName, _, _, _          |
| Pcdata, _, _, _ | `<`                      | `&lt;` | _, _, _, _              |
\end{verbatim}
\hrule
\medskip
\begin{verbatim}
| EnterAttr, _, Url, _ |  |  | Attr, _, _, _; start(Url, htmlCodec) |
\end{verbatim}
\hrule
\medskip
\begin{verbatim}
| CName, _, _, _ | `"[^"]*"?` |  | _, _, _, _ | W: HTML attribute in close tag |
\end{verbatim}
\caption{Representative HTML automaton transition rules.  Each row
  matches on the current context fields (column~1), a regular
  expression or interpolation indicator (column~2), an optional
  substitution (column~3), and specifies the successor context and
  any subsidiary automaton adjustments (column~4).  An optional fifth
  column carries a diagnostic message.  From top to bottom: (1)~a
  less-than followed by a slash and a name character begins a close
  tag; (2--3)~\texttt{<script>} and \texttt{<style>} tags enter
  \texttt{BeforeAttr} with the element type remembered; (4)~a
  less-than followed by a name character begins an open tag; (5)~any
  other less-than is replaced by \texttt{\&lt;} to reduce attack
  surface.  The sixth rule spins up a \texttt{Url} subsidiary
  automaton, bridged via \texttt{htmlCodec}, upon entering a URL-typed
  attribute.  The final rule flags a quoted string inside a close tag
  as a likely error.}
\Description[Markdown tables]{Selected rows from three different Markdown tables containing structured content. Details in caption.}
\label{fig:transition-rules}
\end{figure*}

Subsidiary automata handle processing of nested content languages.  Upon encountering \verb|<a href=|, the HTML automaton spins up a URL parsing automaton and mediate its input via an HTML attribute codec that decodes content as seen by the HTML automaton and re-encodes any modifications made by the URL automaton.  Upon entering a \verb|<style>| element, the HTML automaton spins up a CSS automaton, which may itself spin up a URL automaton upon entering a CSS ``\texttt{url(}'' construct, but the HTML automaton retains responsibility for identifying the ``\verb|</style|'' which ends the CSS body content.  The subsidiary relationship is explicitly represented alongside the context value, so the analysis can track it.

\subsection{Pure Functional Specification of Library Semantics}

The accumulator interface and the transition table together form a complete, purely functional specification of the library's runtime semantics.  The transition function maps a context and an input (either a fixed chunk or an untrusted value) to a pair of an output (the safely encoded content to emit) and a successor context.  Because this specification is expressed in a side-effect-free subset of the host language, it can serve simultaneously as the executable runtime implementation and as the subject of static analysis.

A pure function from contexts to escaping functions provides the bridge: given the current context, it returns the escaper that should be applied to the next interpolated expression result, and the context to adopt after that result has been appended.  The full composition pipeline is therefore a fold over the sequence of fixed and untrusted parts, driven by a pure state-transition function. This observation will be central to the compile-time analysis described in Section~\ref{sec:static-analysis}.  Additionally, from an escaper, the compiler needs a way to express a pure expression that retrieves the escaper at runtime.

\subsection{Reified Security Guarantees and Context-Sensitive Escaping}
\label{subsec:trusted-types}

A tagged expression need not return a string.  It can return a value that reifies the security guarantee that the tag library establishes.  If a value of type \texttt{SafeHtml}, for instance, is unforgeable without going through a library that safely constructs HTML, it attests that the the value's string content is safe to embed in an HTML document.  This design follows the \emph{trademark} pattern described by Morris\thinspace\cite{morris-1973}; a tagged string expression serves as a trusted factory for privileged values.  Security auditing can then focus on the code that constructs those values over the code that those values pass through.

The compositional nature of safe types is illustrated in this example:

\begin{verbatim}
let love = html"I <3 <b>you</b>";

html"""
  "<i id=${love}
  ">${love}</i>
.text ==
"""
  "<i id='I &lt;3 &lt;b&gt;you&lt;/b&gt;'
  ">I &lt;3 <b>you</b></i>
\end{verbatim}

Three aspects of this example deserve attention.  First, the bare \verb|<3| that does not begin a tag is converted to \texttt{\&lt;3} by the automaton's attack-surface-reduction rule; the output language simply doesn't allow the ``<'' character in strings that specify regular text nodes.  Second, when \texttt{love} is interpolated into an HTML \texttt{Pcdata} context, in the body of the \verb|<i>| element, its \texttt{SafeHtml} type signals to the chosen escaper that it has already been composed correctly and it passes through without further escaping, preserving the \verb|<b>| tags intact.  Third, when \texttt{love} is interpolated into an HTML attribute value context, into the \texttt{id} attribute, its content \emph{is} re-escaped, because the attribute context escaper has to preserve the integrity of the delimiting quotes which is why \verb|<b>| and \verb|</b>| become \texttt{\&lt;b\&gt;} and \texttt{\&lt;/b\&gt;}, respectively.

This context-sensitive treatment of precisely typed values is the practical benefit of the reified guarantee: a \texttt{SafeHtml} value is not blindly trusted everywhere, but trusted \emph{appropriately}, according to the requirements of the context into which it is being embedded.  The fifth-column diagnostic mechanism visible in Figure~\ref{fig:transition-rules} will be revisited in Section~\ref{sec:static-analysis}, where we show how the same transition rules that drive runtime escaping can be lifted to produce compile-time warnings and errors visible to developers and adversarial AI coding agents.

Careful API design can help guide developers towards safe idioms, for example allowing an HTTP response body class to accept \texttt{SafeHtml} but not a plain string. When APIs are designed so that type safety implies safety, the practices developers already use to get their code to pass the type checker contribute to security.

\subsection{Use cases for partial event based parsing}
\label{subsec:event-based-parsing-use-cases}

Event-based parsing\thinspace\cite{vadim-2019} is a parsing technique where, instead of producing a syntax tree directly, a parser produces a stream of ``events'' that may correspond to parts of a syntax tree. For example, an event based parser for the JSON text \texttt{\{ "nums": [123] \}} might produce this sequence of events: start record, start property ``nums'', start array, int 123, end array, end property, end record. Events can be fed to a consumer that, when start and finish events pair properly, builds a valid syntax tree; but other analyses and transforms can be done directly on the stream of events that do not require building an entire tree structure in memory.

The transition rules from section \ref{subsec:transition-tables} include a substitution column which specifies alternative text to append instead of that consumed by the regular expression.  Instead of specifying replacement text, this section can specify zero or more \textit{events} which apply to the collector. Turning chunks of fixed content into events allows for two quite different use cases:

\begin{enumerate}
\item Building a syntax tree directly from a quoted expression instead of building text and later parsing it, for example via a JSON tree producer that consumes events like those listed above.
\item Localization of HTML and other semi-structured content languages that need to be presented to human readers in their preferred natural language.
\end{enumerate}

The first use case enables familiar syntax for JSON tree building, \texttt{jsontree`\{ "foo": \$\{bar\} \}`}, which conveniently allows embedded conditionals in the multiline form. It can use events for most tree building steps, but must still append content to a buffer where quoted JSON string nodes contain interpolations.

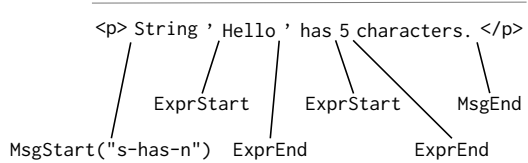
\begin{figure}[htbp]
\centering
\begin{tikzpicture}[
  font=\ttfamily\small,
  >=Stealth,
  line width=0.6pt,
  every node/.style={inner sep=1pt},
]

\node (s0)  at (0,0)   {<p>};
\node (s1)  [right=0pt of s0]  {String };
\node (s2)  [right=0pt of s1]  {'};
\node (s3)  [right=0pt of s2]  {Hello};
\node (s4)  [right=0pt of s3]  {'};
\node (s5)  [right=0pt of s4]  { has };
\node (s6)  [right=0pt of s5]  {5};
\node (s7)  [right=0pt of s6]  { characters.};
\node (s8)  [right=0pt of s7]  {</p>};

\draw[gray, thin]
  ([yshift=6pt]s0.north west) -- ([yshift=6pt]s8.north east);

\node (MsgStart)   at (0,    -1.6) {\textit{MsgStart("s-has-n")}};
\node (ExprStart1) at (1.2,  -1.0) {\textit{ExprStart}};
\node (ExprEnd1)   at (2.1,  -1.6) {\textit{ExprEnd}};
\node (ExprStart2) at (3.2,  -1.0) {\textit{ExprStart}};
\node (ExprEnd2)   at (4.5,  -1.6) {\textit{ExprEnd}};
\node (MsgEnd)     at (5.0,  -1.0) {\textit{MsgEnd}};

\draw (MsgStart.north)   -- (s1.south west);
\draw (ExprStart1.north) -- (s3.south west);
\draw (ExprEnd1.north)   -- (s3.south east);
\draw (ExprStart2.north) -- (s6.south west);
\draw (ExprEnd2.north)   -- (s6.south east);
\draw (MsgEnd.north)     -- (s8.south west);

\end{tikzpicture}
\caption{HTML buffer string with internationalization metadata marks}
\Description[String with marks indexing into it]{A string of HTML across the top reads <p>String 'Hello' has 5 characters</p>. Along the bottom are marks with lines pointing between characters of the string.  MsgStart and MsgEnd point to the end of the <p> and beginning of </p> respectively.  A ExprStart and ExprEnd pair point at the edges of Hello, and another pair point at the beginning and end of 5.}
\label{fig:i18nmarks}
\end{figure}

To understand the second use case, consider HTML corresponding to the ``Zeichenkette'' example\thinspace\cite{gettext-manual} of software localization. \texttt{html\allowbreak "<p><message\allowbreak  i18n="@@s-has-n">\allowbreak String \allowbreak '\$\{s\}' \allowbreak has \$\{n\} \allowbreak characters.\allowbreak </message>\allowbreak </p>"} has \texttt{<message>} tags which mark the beginning and end of translatable substrings, strings that should be sent to a human language translator, so that when rendering HTML for a German speaker, we can produce ``5 Zeichen lang ist die Zeichenkette 'Hello'.'' instead of ``String 'Hello' has 5 characters.'' even though the former swaps the order of the interpolations. To that end, the collector for the HTML tag also collects internationalization (I18N) marks which include the kind of event and its offset into the accumulated content. The marks correspond to the beginning and end of messages, but also to the beginning and end of interpolations into messages. Figure~\ref{fig:i18nmarks} shows the accumulated buffer containing the reference text and the marks that index into it. Together, those allow separating the interpolations out and resituating them in template strings for other human languages.

Event based parsing allows streaming parsing which meshes well with producing trees from familiar syntax tailored by embedded flow control. But it also extends to extracting and acting on content-language specific metadata that need to be acted upon and erased by the content producer before the content reaches the parser. With ahead of time (section \ref{sec:static-analysis}) parsing, or runtime memoization of transitions, the cost of converting text into events is amortized over many executions.

\section{Performing Analysis Ahead of Time}
\label{sec:static-analysis}

The compile-time optimization described in this section applies to accumulator implementations that satisfy a set of structural conventions.  We state these preconditions first, then illustrate the analysis with a running example.

\subsection{Preconditions}

\paragraph{Inlinable accumulator construction.}
The compiler must be able to simulate the expression that creates a fresh accumulator context, and a collector. It suffices for the tag's \texttt{newAccumulator()} (see Figure~\ref{fig:tagged-desugar}) entry point to delegate to a pure static method on the accumulator type whose body simply constructs a known, non-extensible type which starts with a known zero context, and which allocates a collector of known type.

\paragraph{State confined to context, subsidiary, and collector.}
The accumulator's only mutable state must be the current context, the current subsidiary automaton (if any), and its \emph{collector}: a mutable value that can receive fixed string parts (with any substitutions applied) and escaped interpolation results.  All state transitions are delegated to pure functions, typically but not necessarily implemented using the transition tables described in Section~\ref{subsec:transition-tables}, that the compiler can evaluate symbolically or perform in an interpreter.  The optimization goal is to \emph{erase} the accumulator entirely, replacing calls that manipulate it with calls that interact directly with the collector, and rewriting each \texttt{appendUnsafe} call as an \texttt{append} to the collector that receives the result of applying the escapers chosen based on context to the untrusted expression.

\paragraph{A merge operator for rejoining control flow.}
When control flow paths diverge and later rejoin, e.g. at the end of an \texttt{if}--\texttt{else}, or at the continue and break edges of a loop, the compiler needs a context that is valid regardless of which path was taken.  The accumulator type provides a pure \emph{merge operator} that takes two or more contexts and returns a context that conservatively over-approximates both.  Merges need not be perfectly semantics-preserving; they may be \emph{fail-stop}, meaning the merge is permitted to return an error context accompanied by a diagnostic message severe enough to indicate that the resulting compilation output should not be used in production.  Loops require the analysis to iterate to a fixed point: the merge of the context at the loop entry with the context produced at the end of each loop body must eventually stabilize.

\paragraph{An optional end-context validity check.}
It is useful, though not required, for the accumulator type to expose a predicate that declares whether a context is a valid \emph{terminal} context.  After propagating contexts through the entire control flow graph of the desugared expression, the compiler evaluates this predicate on the final merged context and emits any appropriate diagnostics.  For example, the expression

\begin{verbatim}
    html"<a href=hello>link</a"
\end{verbatim}

\noindent
never triggers a bad transition, but it does not terminate in a context corresponding to a well-formed HTML fragment: the \texttt{</a>} end tag is missing a final '>'.  The end-context check surfaces this as a compile-time warning rather than a silent runtime hazard. Not all languages allow for concatenation (there is no general way to concatenate two URLs and get their combined semantics), but where a language is, end state checks help produce fragments that preserve semantics under concatenation.

\paragraph{Inlinable result construction.}
The compiler must be able to construct the final value for the string expression without going through the full accumulator.  It suffices for \texttt{accumulator.collected()} to delegate to a pure function that operates only on the collector, allowing the compiler to substitute \texttt{tag\allowbreak .fromCollector\allowbreak (collector)} once the accumulator has been erased.

\subsection{Running Example}

Figure~\ref{fig:running-example-source} shows a tagged block expression that generates an HTML list used to explain the transformation pipeline.

\begin{figure}[htbp]
\begin{verbatim}
html"""
  "<ul>
  :for (let item of items) {
    "  <li><a href=${item.url}>${item.label}</a></li>
  :}
  "</ul>
\end{verbatim}
\caption{A tagged block expression generating an HTML unordered list.
  Each item contributes a hyperlink whose target and label are drawn
  from untrusted values.}
\Description[A tagged string expression]{A string expression using the tag ``html'' and which contains an embedded for loop.  The first and last textual lines open and close an HTML unordered list and the loop body contributes one list item HTML element for each iteration.  Inside the list item are interpolated expressions for a link URL and text.}
\label{fig:running-example-source}
\end{figure}

\subsection{Desugaring}

The compiler first desugars the tagged block expression into the accumulator-based form shown in Figure~\ref{fig:running-example-desugar}, following the mechanical translation described in Section~\ref{sec:string-composition-syntax}.

\begin{figure}[htbp]
\small
\begin{verbatim}
do {
  let accum = html.newAccumulator();
  accumulator.appendFixed("<ul>\n");
  for (let item of items) {
    accum.appendFixed("  <li><a href=");
    accum.appendUnsafe(item.url);
    accum.appendFixed(">");
    accum.appendUnsafe(item.label);
    accum.appendFixed("</a></li>\n");
  }
  accum.appendFixed("</ul>\n");
  accum.collected()
}
\end{verbatim}
\caption{Desugared accumulator-based form of Figure~\ref{fig:running-example-source}.}
\Description[The syntactic desugaring of the previous figure]{A block of statements that creates a new accumulator and feeds it string literals via its appendFixed method and interpolated expressions via its appendUnsafe method}
\label{fig:running-example-desugar}
\end{figure}

\subsection{Flow-Sensitive Context Propagation}

Before erasing the accumulator, the compiler performs a \emph{flow-sensitive} analysis, propagating contexts forward through each to \texttt{append\allowbreak Fixed} or \texttt{append\allowbreak Unsafe}.  It applies the merge operator at loop edges and at join points after conditionals, iterating to a fixed point.  If the \texttt{html} context-propagation function accepts an optional diagnostic callback, the optimization pass supplies one that forwards messages to the compiler's error and warning reporter, prepending each message with the source position of the \texttt{append\allowbreak Fixed} or \texttt{append\allowbreak Unsafe} call argument.

The result of the analysis is most easily understood by imagining the
compiler inserting assertions and notes before and after each append, as shown
in Figure~\ref{fig:running-example-annotated}.

\begin{figure}[htbp]
\small
\begin{verbatim}
do {
  let accum = html.newAccumulator();
  // context: (Pcdata, _, _, _)
  accum.appendFixed("<ul>\n");
  // context: (Pcdata, _, _, _)
  for (let item of items) {
    accum.appendFixed("  <li><a href=");
    // epsilon transition inserts '"'
    // context: (BeforeValue, _, Url, _)
    // esc [HtmlAttributeEscaper,
    //      UrlPrefixFilteringEscaper]
    accum.appendUnsafe(item.url);
    // epsilon transition inserts '"'
    // context: (AfterValue, _, _, _)
    accum.appendFixed(">");
    // context: (Pcdata, _, _, _)
    // esc [HtmlPcdataEscaper]
    accum.appendUnsafe(item.label);
    // context: (Pcdata, _, _, _)
    accum.appendFixed("</a></li>\n");
    // context: (Pcdata, _, _, _)
  }
  // merged context: (Pcdata, _, _, _)
  accum.appendFixed("</ul>\n");
  // context: (Pcdata,_,_,_) [ok at end]
  accum.collected()
}
\end{verbatim}
\caption{Annotated desugaring showing the context at each program
  point.  Epsilon transitions on the \texttt{href} interpolation site
  cause the automaton to insert quotation marks automatically.  The
  \texttt{item.url} interpolation site selects two escapers in
  composition: one from the HTML automaton and one from the URL
  subsidiary automaton spun up upon entering the \texttt{href}
  attribute.  The final context is a valid terminal context, so no
  end-context diagnostic is issued.}
\Description[The same code as the previous figure but with comments]{Comments between the statements from the previous figure describe the parse context as a tuple of enum values.}
\label{fig:running-example-annotated}
\end{figure}

The analysis observes that after the loop body, the context is the same \texttt{Pcdata} context as before the loop; the fixed point is reached in a single iteration.  The end-context check confirms that \texttt{Pcdata} is a valid terminal context for a \texttt{SafeHtml} value, so no diagnostic is issued.

The diagnostics produced by this analysis are not only useful to human developers.  Iterative AI code generation techniques such as Recursive Criticism and Improvement (RCI)\thinspace\cite{kim2023rci} work by prompting a model to identify problems in its own output and then revise it. The effectiveness of such techniques is bounded by the quality of the feedback available: a model asked to review its own code for injection vulnerabilities must rely on internalized heuristics, which are imperfect. Compile-time diagnostics generated by the static analysis described here provide an objective, position-keyed signal (\emph{the surrounding content could lead to two possible interpretations of an interpolation}) that an RCI loop can include verbatim in its critique prompt, grounding the model's self-revision in a verifiable fact rather than a heuristic judgment.  The same diagnostics, surfaced in an IDE, serve an equivalent function for human developers working with AI coding assistants.

\subsection{Accumulator Erasure}
\label{subsec:accumulator-erasure}

Once the analysis is complete and no blocking diagnostics have been issued, the compiler erases the accumulator entirely, producing an internal representation equivalent to the code shown in Figure~\ref{fig:running-example-optimized}.

\begin{figure}[htbp]
\small
\begin{verbatim}
do {
  let collector: StringBuilder =
    html.newCollector();
  collector.append("<ul>\n");
  for (let item of items) {
    collector.append("  <li><a href=\"");
    collector.append(
      HtmlAttributeEscaper.instance.apply(
        UrlPrefixFilteringEscaper.instance
        .apply(item.url)
      )
    );
    collector.append("\">");
    collector.append(
      HtmlPcdataEscaper.instance
      .apply(item.label)
    );
    collector.append("</a></li>\n");
  }
  collector.append("</ul>\n");
  html.fromCollector(collector)
}
\end{verbatim}
\caption{Optimized output after accumulator erasure.  The accumulator
  and its context fields have been eliminated; each former
  \texttt{appendUnsafe} call is rewritten as an \texttt{append} of
  the escaped value, with the escaper chain determined statically by
  the flow-sensitive analysis.  Quotation marks elided from the
  source are inserted by the rewriter at the sites identified by the
  epsilon-transition substitution rules.}
\Description[Code that performs steps similar to the previous figure but refocused]{The steps from the previous figure are done the same, but appendFixed are now just appends to a StringBuilder, and appendUnsafe steps are now regular append but with explicit calls to escaper and filter functions.}
\label{fig:running-example-optimized}
\end{figure}

\subsection{Runtime Cost}
\label{subsec:runtimecost}

The optimized output in Figure~\ref{fig:running-example-optimized} makes the residual runtime cost explicit.  The accumulator, context values, and subsidiary automaton bookkeeping have been entirely eliminated at compile time.  The only overhead relative to na\"{i}ve string concatenation is the cost of the escaper functions themselves.  Each of these is a simple value transform that can be hand optimized. Crucially, this residual cost is not an artifact of the security scheme. It is the \emph{irreducible cost of security at runtime}.  Any correct content composition, however written, must apply equivalent transforms to untrusted values.

As discussed in section \ref{subsec:event-based-parsing-use-cases}, the latest implemention of the HTML tag does not use \textit{StringBuilder} alone. It uses a thin wrapper around StringBuilder that also tracks a list of I18N marks. Other tags do reduce fully as above, and the HTML tag does benefit from compile-time parsing of I18N metadata that also enables auto-extraction of message bundles for human-language translators.

\section{Conclusions}
\label{sec:conclusions}

Modern distributed systems are multi-language systems comprising programs that send and receive messages in complex content languages.  These messages, which receivers trust, are composed of values of unknown and practically unknowable provenance.

Libraries allow sanitizing and DSLs allow for composing content, but neither have effectively allowed general purpose programming languages' tools to step outside their box and surface analyses of content languages in time for developers and code auditors to adjust.

To build reliable systems, we need to integrate the work of different specialties: developers who understand their users, security engineers who understand how to produce defensive code and who keep up-to-date with new attack techniques, code testers and auditors who can catch mistakes, and programming language designers who produce tools that integrate their efforts.

This paper contributes:

\begin{enumerate}
\item A direction for evolving quoted expression syntax that avoids
  the trap of syntactic sugar abandoned as requirements rise in complexity,
  enabling both effective review and extensible, secure semantics for composition.
\item A desugaring of that syntax into a cooperation between an
  \emph{accumulator}, which makes security-relevant decisions, and a
  \emph{collector}, which assembles output-language fragments.
\item Compiler techniques with low meta-programming requirements that enable
  ahead-of-time analysis of embedded content-language fragments, erasing
  accumulators after inlining their security decisions so that runtime
  cost depends only on collection, and surfacing analyses of content-language
  templates as compiler diagnostics.
\item \emph{Transition tables}: artifacts maintainable by security
  engineers once per content language, shareable across many
  programming language communities, and amenable to shared testing and
  verification infrastructure such as grammar fuzzers.
\end{enumerate}

Our hope is that, taken together these contributions will relieve generalist software developers of the burden of secure composition via ad-hoc escaping and filtering. This is timely because generative AI systems are showing persistent, high rates of failure on security related tasks\thinspace\cite{veracode2026} including many classes of vulnerabilities that center around content composition mistakes\thinspace\cite{negriribalta2024}. Bringing security engineers expertise to bear to make the path of least resistance in coding also a secure way, and reporting early and loudly about flaws in content templates we hope will help both human developers and toolchains that incorporate AI code generators and auditors.

We hope this will also simplify the economics of security work. As the way software is developed changes, the security community will see new instances of composition vulnerabilities emerge. A general framework for safe composition gives security engineers a pathway to address them with the same tools and idioms, rather than requiring a new domain-specific solution each time. Sharing development and maintenance of artifacts, like transition tables, across many general-purpose programming language ecosystems also allows the security engineering community to support everyone, instead of working to secure only widely used languages just in time for them to become legacy programming languages.

\section{Related Work}
\label{sec:related}

The earliest deployed solution to automatic context-aware escaping was introduced by Boutros~et~al.\thinspace\cite{boutros2009autoescape}, who added auto-escaping to Google-developed general-purpose template systems using dynamic analysis to infer the syntactic context of each expression site at render time.  This approach demonstrated dramatically reduced XSS vulnerability rates in large-scale production systems but is confined to the specific template languages in which it was implemented.

Authors' prior work\thinspace\cite{samuel2011csas} subsequently formalized context-sensitive sanitization for web template languages using a type-qualifier system, providing static guarantees in place of the runtime inference used by earlier dynamic approaches.  Kern~et~al.\thinspace\cite{kern2014} extended static enforcement further, demonstrating that whole-program analysis over a domain-specific template language can eliminate an entire class of injection vulnerabilities at compile time.

These prior systems share a common limitation: each is specific to a particular template and content language and requires dedicated build-system integration.  Google's Secure by Design report\thinspace\cite{google2024securebydesign} documents the practical security gains achievable when such approaches are applied at scale\thinspace\cite[p.~{7}]{google2024securebydesign}, but also illustrates that adoption remains bounded by the friction of adopting a separate toolchain.  The present work addresses this limitation by embedding secure composition primitives directly into general-purpose programming language syntax and type systems, eliminating the need for a separate template language altogether.

A different approach to avoiding a separate DSL build pipeline is to embed content-language syntax directly into a general-purpose programming language.  ECMAScript for XML (E4X), standardized as ECMA-{357}\thinspace\cite{ecma357-2005}, extended JavaScript with native XML literals and expression interpolation inside XML structure, allowing developers to produce XML DOM trees without string concatenation or a separate template processor.  JSX\thinspace\cite{react-jsx}, introduced by Meta for the React framework, similarly allows HTML-like tag syntax inline in JavaScript, eliminating the need to integrate a separate HTML template language into the build pipeline.  Both approaches share the advantage of familiar syntax and direct toolchain integration.  However, their security affordances do not extend naturally to \emph{nested} content languages; when a React component embeds JavaScript event handlers, CSS, or URLs inside HTML, the developer is responsible for knowing and applying the correct escaping for each nested context.  The latency of this limitation is illustrated by the \texttt{javascript:} URL injection vector in JSX. It was not until React~{16.9} (released 2019) that the framework began warning developers against passing dynamic values into \texttt{href} attributes that could resolve to \texttt{javascript:} URLs\thinspace\cite{react-16-9-changelog}, and not until React~{19} that such URLs were fully blocked\thinspace\cite{react-changelog}.

A third approach retains standard JavaScript syntax by exploiting \emph{tagged template literals}, introduced in ECMAScript~{2015}.  A tag function receives the interleaved static string parts and dynamic expression values of a template literal as separate arguments, enabling a library to apply context-sensitive escaping at the interpolation sites.  Python's PEP-750\thinspace\cite{pep750} introduces \texttt{t"..."} strings which are intended to preserve the same distinction for security use cases including SQL injection and XSS. Security libraries taking this approach include the \texttt{html} tag in lit-html\thinspace\cite{lit-html-npm}, which uses the static/dynamic split to protect against XSS by treating expression values as text content rather than raw markup, and \texttt{safesql}\thinspace\cite{safesql-npm}, which applies MySQL and PostgreSQL context-aware escaping to prevent SQL injection.  The fundamental limitation of this approach is that the analysis occurs entirely at runtime: the tag function processes the template only when it is evaluated, so no compile-time diagnostics are possible.  Additionally, tagged template literals do not support general control flow (conditionals, loops) within the template itself, limiting their ability to extend to complex use cases.

The pattern of using types like \textit{SafeHtml}, which attest to the trustworthiness of their values for some specific use, aligns \textit{type safety} checks (whether compiler enforced or based on runtime type information) with \textit{safety} against a class of attack. It is used in the W3C Trusted Types specification\thinspace\cite{w3c-trusted-types} which privileges policy-created values like \textit{TrustedHTML} values over plain strings of unknown provenance at security sensitive \textit{sinks} like \texttt{HTMLElement}'s setter for \texttt{innerHTML}.

\bibliographystyle{ACM-Reference-Format}
\bibliography{bibfile}

@misc{cwe-top25-2025,
  author       = {{MITRE Corporation}},
  title        = {{2025 CWE Top 25 Most Dangerous Software Weaknesses}},
  year         = {2025},
  howpublished = {\url{https://cwe.mitre.org/top25/archive/2025/2025_cwe_top25.html}},
  note         = {Accessed: 2026-03-19}
}

@inproceedings{kim2023rci,
  author    = {Geunwoo Kim and Pierre-Louis Poirion and Minsu Park and
               Dong-Gi Lee and Byungkwon Choi and Donghyun Kang and
               Jiyong Jang},
  title     = {Language Models can Solve Computer Tasks},
  booktitle = {Advances in Neural Information Processing Systems},
  series    = {NeurIPS~'23},
  year      = {2023},
  url       = {https://arxiv.org/abs/2303.17491}
}

@misc{boutros2009autoescape,
  author       = {Jad S. Boutros},
  title        = {Reducing \{XSS\} by Way of Automatic Context-Aware Escaping in Template Systems},
  year         = {2009},
  month        = mar,
  howpublished = {Google Online Security Blog, \url{https://security.googleblog.com/2009/03/reducing-xss-by-way-of-automatic.html}},
  note         = {Accessed: 2026-03-19}
}

@article{kern2014,
  title = {Securing the Tangled Web},
  author = {Christoph Kern},
  year = {2014},
  URL = {http://dx.doi.org/10.1145/2643134},
  journal = {Communications of the ACM},
  pages = {38--47},
  volume = {57, no. 9}
}

@inproceedings{samuel2011csas,
  author    = {{Anonymous Authors}},
  title     = {{Self-citation omitted for double-blind review}},
  booktitle = {Proceedings of ...},
  year      = {2011},
  note      = {Details omitted for double-blind review}
}

@techreport{google2024securebydesign,
  author      = {{Google}},
  title       = {An Overview of {Google}'s Commitment to Secure by Design},
  institution = {Google},
  year        = {2024},
  month       = oct,
  type        = {White Paper},
  url         = {https://static.googleusercontent.com/media/publicpolicy.google/en//resources/google_commitment_secure_by_design_overview.pdf},
  note        = {Accessed: 2026-03-19}
}

@techreport{ecma357-2005,
  author      = {{Ecma International}},
  title       = {ECMAScript for XML (E4X) Specification},
  institution = {Ecma International},
  number      = {ECMA-357},
  edition     = {2nd},
  year        = {2005},
  month       = dec,
  url         = {https://ecma-international.org/publications-and-standards/standards/ecma-357/},
  note        = {Withdrawn 2021}
}

@misc{react-jsx,
  author       = {{Meta Platforms, Inc.}},
  title        = {{JSX} in {React} -- Introducing {JSX}},
  howpublished = {\url{https://react.dev/learn/writing-markup-with-jsx}},
  note         = {Accessed: 2026-03-19}
}

@misc{react-16-9-changelog,
  author       = {{Meta Platforms, Inc.}},
  title        = {React v16.9.0 and the Roadmap Update},
  year         = {2019},
  month        = aug,
  howpublished = {React Blog,
                  \url{https://legacy.reactjs.org/blog/2019/08/08/react-v16.9.0.html}},
  note         = {Accessed: 2026-03-19}
}

@misc{react-changelog,
  author       = {{Meta Platforms, Inc.}},
  title        = {{React} {CHANGELOG}},
  year         = {2025},
  howpublished = {\url{https://github.com/facebook/react/blob/main/CHANGELOG.md}},
  note         = {Accessed: 2026-03-19}
}

@misc{lit-html-npm,
  author       = {{Google LLC}},
  title        = {lit-html},
  year         = {2025},
  howpublished = {npm, \url{https://www.npmjs.com/package/lit-html}},
  note         = {Accessed: 2026-03-19}
}

@misc{safesql-npm,
  author       = {Anonymous Authors},
  title        = {Self-citation omitted for double-blind review},
  year         = {2019},
  howpublished = {anonymized, \url{https://example.com/anonymized}},
  note         = {Accessed: 2026-03-19}
}

@misc{pep498,
  author       = {Eric V. Smith},
  title        = {{PEP} 498 -- Literal String Interpolation},
  year         = {2015},
  howpublished = {Python Enhancement Proposals,
                  \url{https://peps.python.org/pep-0498/}},
  note         = {Accepted 2016-08-08. Accessed: 2026-03-19}
}

@misc{pep750,
  author       = {Jim Baker},
  title        = {{PEP} 750 -- Template Strings},
  year         = {2024},
  howpublished = {Python Enhancement Proposals,
                  \url{https://peps.python.org/pep-0750/}},
  note         = {Accepted 2025-04-10. Accessed: 2026-04-23}
}

@inproceedings{finifter2008joe-e,
  author    = {M. Finifter and A. Mettler and N. Sastry and D. Wagner},
  title     = {Verifiable functional purity in Java},
  booktitle = {15th ACM Conference on Computer and Communications Security},
  series    = {CCS'08},
  year      = {2008},
  pages     = {161--175},
  url       = {https://people.eecs.berkeley.edu/~daw/papers/pure-ccs08.pdf}
}

@mastersthesis{tarvankrieken-2023,
  author = {Tar van Krieken},
  title  = {Deriving Syntax Highlighting Grammars from Character-Level
            Context-Free Grammars: Algorithm Development, Analysis, and
            Future Directions},
  school = {Eindhoven University of Technology},
  year   = {2023},
  month  = dec,
  url    = {https://homepages.cwi.nl/~jurgenv/theses/TarVanKrieken.pdf},
  note   = {Accessed: 2026-03-19}
}

@article{morris-1973,
  author  = {James H. {Morris, Jr.}},
  title   = {Protection in Programming Languages},
  journal = {Communications of the {ACM}},
  volume  = {16},
  number  = {1},
  pages   = {15--21},
  year    = {1973},
  month   = jan,
  doi     = {10.1145/361932.361937}
}

@techreport{w3c-trusted-types,
  author      = {Krzysztof Kotowicz and Mike West},
  title       = {Trusted Types},
  institution = {World Wide Web Consortium ({W3C})},
  year        = {2024},
  type        = {{W3C} Working Draft},
  url         = {https://www.w3.org/TR/trusted-types/},
  note        = {Accessed: 2026-03-19}
}

@misc{jep378,
  author       = {Jim Laskey},
  title        = {{JEP} 378: Text Blocks},
  year         = {2020},
  howpublished = {{OpenJDK} Java Enhancement Proposal,
                  \url{https://openjdk.org/jeps/378}},
  note         = {Finalized in {JDK}~15. Accessed: 2026-03-19}
}

@book{raymond-taoup-2003,
  author    = {Eric S. Raymond},
  title     = {The Art of {Unix} Programming},
  publisher = {Addison-Wesley},
  year      = {2003},
  isbn      = {0-13-142901-9},
  note      = {Rule of Least Surprise: \url{http://www.catb.org/~esr/writings/taoup/html/ch01s06.html}}
}

@inproceedings{vadim-2019,
author = {Zaytsev, Vadim},
title = {Event-based parsing},
  year = {2019},
  isbn = {9781450369862},
  publisher = {Association for Computing Machinery},
  address = {New York, NY, USA},
  url = {https://doi.org/10.1145/3358503.3361275},
  doi = {10.1145/3358503.3361275},
  booktitle = {Proceedings of the 6th ACM SIGPLAN International Workshop on Reactive and Event-Based Languages and Systems},
  pages = {31–40},
  numpages = {10},
  location = {Athens, Greece},
  series = {REBLS 2019}
}

@manual{gettext-manual,
  author       = {Drepper, Ulrich and Meyering, Jim and Pinard, Fran\c{c}ois
                  and Haible, Bruno},
  title        = {{GNU} gettext utilities},
  organization = {Free Software Foundation},
  edition      = {0.26},
  year         = {2026},
  url          = {https://www.gnu.org/software/gettext/manual/gettext.html#Special-Comments-preceding-Keywords},
  note         = {Section: Special Comments preceding Keywords}
}

@techreport{veracode2026,
  author      = {{Tischler, Natalie and Ariganello, Joe}},
  title       = {{2026 State of Software Security: Prioritize, Protect, Prove}},
  institution = {Veracode},
  year        = {2026},
  url         = {https://www.veracode.com/resources/state-of-software-security},
  note        = {Data analysis by David Severski and Wade Baker (Cyentia Institute)},
}

@ARTICLE{negriribalta2024,
  AUTHOR={Negri-Ribalta, Claudia  and Geraud-Stewart, Rémi  and Sergeeva, Anastasia  and Lenzini, Gabriele},
  TITLE={A systematic literature review on the impact of AI models on the security of code generation},
  JOURNAL={Frontiers in Big Data},
  VOLUME={Volume 7 - 2024},
  YEAR={2024},
  URL={https://www.frontiersin.org/journals/big-data/articles/10.3389/fdata.2024.1386720},
  DOI={10.3389/fdata.2024.1386720},
  ISSN={2624-909X}}

\appendix 

\section{Open Science} 
\label{sec:openscience}

The proof of concept of these ideas has been implemented in the Temper programming language. Its compiler and command line tools are available under open source licenses at \texttt{temperlang.dev/}, and includes the optimization passes.

The security engineering library \texttt{secure-composition}, which implements secure composition tags for web-platform and database languages is currently closed source; a complete but unlicensed snapshot is available at \textit{https://anonymous\allowbreak .4open\allowbreak .science\allowbreak /r/seccomp-snapshot-3B7C\allowbreak /README.md}. Its authors worry that releasing it too early would complicate backwards-compatibility breaking changes that might be necessary for user security until more pilots have completed.

Once the Temper compiler and the proof of concept library are locally available, from the directory which a researcher \texttt{git clone}d the library into, they may run \texttt{temper repl -w . -e 'let \{...\} = import("secure-composition/html")'} to start the language's interactive playground with the web security portion of the library available.  Similarly, \texttt{temper test -b js} will run the embedded testcases allowing a researcher to add their own tests and check their results.

\section{Ethical Considerations} 
\label{sec:ethicalconsiderations}

This work involves computer security. Failures in it can put humans at risk. It is focused on new defensive measures, discussion of which advantages defenders over attackers. Testing defensive techniques involve acting as if attacking but no people were put at risk during the development because no systems with user data or non-using humans' data were involved in testing. Due to the nature of the work, there was no need to attack any existing system or to simulate a system end-to-end to attack that might use real data. All simulated attacks were done by implementors writing unit tests, using the programming language's interactive read-eval-print loop (REPL) with inputs they personally supplied, and writing small programs that were never deployed to an environment with production databases.

Incorrectness in security mitigations can put people at risk. Without a standardized approach to content composition, every ad-hoc decision is a potential security failure. With a standardized approach, the weaknesses can be systematically tested, enumerated, and potentially addressed.  As such, this paper falls within the tradition of openly discussing defensive mechanisms' potential and flaws so that early deployments can be carefully combined with other mitigations that provide defense in depth, and later deployments build upon a foundation with a quantifiable record of efficacy.

\section{Generative AI Usage}
\label{sec:genaiusage}

The main use of generative AI in this work was to fit LaTeX source into the required template (checked by humans using \texttt{git diff}) and debug \texttt{pdflatex} errors. Generative AI was used to generate appropriate bibliography entries from human-provided URLs which were then human reviewed, to copyedit and reword human-authored content for a few sections. In the proof of concept implementation described, generative AI use has been limited to finding inconsistencies in pre-existing code, to find opportunities for spot optimizations in the source code, and to suggest additional test cases.

\end{document}